\documentclass{aa}  

\usepackage{graphicx}
\usepackage{xcolor}
\usepackage{txfonts}
\begin{document}

   \title{Whistler waves observed by Solar Orbiter / RPW between 0.5 AU and 1 AU }


   \author{M. Kretzschmar \inst{1}
\and
T. Chust\inst{2}
\and
V. Krasnoselskikh\inst{1,13}
\and
D. Graham\inst{3}
\and
L. Colomban\inst{1}
\and
M. Maksimovic\inst{5}
\and
Yu. V. Khotyaintsev\inst{3}
\and
J. Soucek\inst{6}
\and
K. Steinvall\inst{3}
\and
O. Santolík\inst{6}
\and
G. Jannet\inst{1}
\and
J.-Y. Brochot\inst{1}
\and
O. Le Contel\inst{2}
\and
A. Vecchio\inst{5,12}
\and
X. Bonnin\inst{5}
\and
S. D. Bale\inst{13, 14, 15}
\and
C. Froment\inst{1}
\and
A. Larosa\inst{1}
\and
M. Bergerard-Timofeeva\inst{1}
\and
P. Fergeau\inst{1}
\and
E. Lorfevre\inst{8}
\and
D. Plettemeier\inst{9}
\and
M. Steller\inst{10}
\and
\v{S}. \v{S}tver\'ak\inst{11}
\and
P. Tr\'avn\'i\v{c}ek\inst{13,11}
\and
A. Vaivads\inst{3,16}
\and
T. S. Horbury\inst{17}
\and
H. O'Brien \inst{17}
\and
V. Evans \inst{17}
\and
V. Angelini \inst{17}
\and
C. Owen \inst{18}
\and
P. Louarn \inst{19}
}

  \institute{LPC2E, UMR7328 CNRS, University of Orléans, 3A avenue de la recherche scientifique, Orléans, France
              \email{matthieu.kretzschmar@cnrs-orleans.fr}
              \and LPP, CNRS, Ecole Polytechnique, Sorbonne Université, Observatoire de Paris, Université Paris-Saclay, Palaiseau, Paris, France 
              \and Swedish Institute of Space Physics (IRF), Uppsala, Sweden
              \and Department of Space and Plasma Physics, School of Electrical
              Engineering and Computer Science, Royal Institute of Technology,
              Stockholm, Sweden
              \and LESIA, Observatoire de Paris, Université PSL, CNRS, Sorbonne
              Université, Université de Paris, Meudon, France
             \and Institute of Atmospheric Physics, Czech Academy of Sciences,
              Prague, Czech Republic
              \and Faculty of Mathematics and Physics, Charles University, Prague, Czech Republic
              \and CNES, 18 Avenue Edouard Belin, 31400 Toulouse, France
              \and Technische Universität Dresden, Würzburger Str. 35, D-01187 Dresden, Germany 
              \and Space Research Institute, Austrian Academy of Sciences, Graz, Austria 
              \and Astronomical Institute of the Czech Academy of Sciences, Prague, Czechia                 \and Radboud Radio Lab, Department of Astrophysics, Radboud University, Nijmegen, The Netherlands
              \and Space Sciences Laboratory, University of California, Berkeley, CA, USA
              \and Physics Department, University of California, CA, USA
              \and Stellar Scientific, Berkeley, CA, USA
 \and Department of Space and Plasma Physics, School of Electrical Engineering and Computer 
Science, Royal Institute of Technology, Stockholm, Sweden
 \and Department of Physics, Imperial College, SW7 2AZ London, UK
              \and Mullard Space Science Laboratory, University College London, Holmbury St. Mary, Dorking, Surrey RH5 6NT, UK
              \and Institut de Recherche en Astrophysique et Planétologie, 9, Avenue du Colonel ROCHE, BP 4346, 31028 Toulouse Cedex 4, France
             }

   \date{Received September 15, 1996; accepted March 16, 1997}

 
  \abstract
   {The solar wind evolution differs from a simple radial expansion and wave-particle interactions are supposed to be the major cause for the observed dynamics of the electron distribution function. In particular, whistler waves are thought to inhibit the electron heat flux and  ensure the diffusion of the field aligned energetic electrons (Strahl) to replenish the halo population.}
   {
   The goal of our study is to detect and characterize the electromagnetic waves that can modify the electron distribution functions, with a special attention to   whistler waves. }
   {We analyse in details the electric and magnetic field fluctuations observed by the Solar Orbiter spacecraft during its first orbit around the Sun between 0.5 and 1 AU. Using data of the Search Coil Magnetometer and electric antenna, both parts of the Radio and Plasma Waves (RPW) instrumental suite, we detect the electromagnetic waves with frequencies above 3 Hz and determine the statistical distribution of their amplitudes, frequencies, polarization and k-vector as a function of distance. We also discuss relevant instrumental issues regarding the phase between the electric and magnetic measurements and the effective length of the electric antenna.}
   {An overwhelming majority of the observed waves are right hand circularly polarized in the solar wind frame and identified as outward propagating and quasi parallel whistler waves. Their occurrence rate increases by a least a factor two from 1 AU to 0.5 AU. These results are consistent with the regulation of the heat flux by the whistler heat flux instability. Near 0.5 AU, whistler waves are found to be more field-aligned and to have smaller normalized frequency ($f/f_{ce}$), larger amplitude, and larger bandwidth than at 1 AU.}
   {}


 \maketitle
%

\section{Introduction}

 The properties of the solar wind are known to change along its propagation in the interplanetary space. 
 Velocity distribution functions (VDF) of ions and electrons are supposed to be far from equilibrium in the source region of the wind, even under quiet conditions, and the observed dynamics of the different constituents of the wind still raises several questions. Our study is associated with the lightest wind constituent, electrons. It is widely accepted that they carry the major part of the heat flux, both in the fast and slow wind, thus their role in the energy balance is very important.\\ Electron VDF consists of four components. Three are isotropic: the thermal core distribution, the energetic halo distribution in the energy range from hundreds of eV to several keV, and the even more energetic population, the superhalo, from several keV to several hundreds keV. The fourth population, the so called Strahl, is quite strongly anisotropic and consists of magnetic field aligned and outward propagating energetic electrons in approximately the same energy range as the halo \citep{Rosenbauer1977,Feldman1978, Pilipp1987,Pilipp1987b}. It is supposed that the heat flux in slow and fast wind is carried by different populations of suprathermal electrons. In the slow wind it is carried by the halo because of its drift velocity shifted in opposite direction in the plasma reference frame, while in the fast wind it is carried by the Strahl population. Suprathermal electrons  are supposed to be created in the tenuous low corona and  their evolution is supposed to be weakly collisional or collisionless. This suggests that their dynamics is mainly determined by the wave particle interactions. \\
  In the fast wind, it has been deduced \citep{2005JGRA..110.9104M} from HELIOS measurements that the strahl population significantly decreases between 0.5 AU and 1 AU while the halo population increases at the same time. Whistler waves are supposed to play an important role in the angular diffusion of the strahl electron, and this is supported by the recent observations of \cite{2021arXiv210106723J} who found an increase of the electrons pitch angle widths in the presence of whistler waves.  Parallel whistler waves, however, can not ensure an effective angular diffusion of the strahl electrons and it has been  proposed \citep[\textit{e.g.,}][]{Vasko_2019} that the diffusion is provided by oblique whistler waves that may be generated when the strahl population has a narrow angular width.

 
 Additionally, the estimates of the heat flux based on measurements of the electron VDF are often found to be significantly smaller than the  Spitzer-Härm \citep{Spitzer:1953p504} collisional heat flux \citep{Feldman76}. This implies that wave particle interactions play an important role in the heat flux inhibition. One of the possible mechanisms that may ensure this is the diffusion of suprathermal halo electrons resulting in the decrease of the halo relative velocity as a result of their interaction with whistler waves. It was shown by several authors \citep{Gary1975,1994JGR....9923391G,Feldman76} that in the slow wind the heat flux instability can create quasi-parallel whistler waves that may scatter the suprathermal electrons and therefore regulate the heat flux. Such a mechanism is supported by several observations. Using CLUSTER data, \cite{0004-637X-796-1-5} found in about 10\% of the analysed spectra the presence of field aligned, narrow band, right handed and circularly polarized waves that they interpreted as whistlers, and that their presence was favored by a larger heat flux.  \cite{2019ApJ...878...41T} and \cite{2019ApJ...870L...6T} presented observations consistent with the whistler heat flux instability (WHFI) producing quasiparallel whistlers. However, \cite{2020PhPl...27h2902V} also suggested that the WHFI cannot efficiently regulate the electron heat flux and that its reduction could be attributed to anti parallel whistlers produced by some other instability (for example, whistler temperature anisotropy instability (WTAI)). It is more difficult to detect these anti parallel whistlers because they have lower frequencies and often smaller amplitude than parallel whistlers. \cite{Agapitov_2020} found during the first perihelion of Parker Solar Probe the presence of numerous sunward whistlers whose propagation direction relative to the background magnetic field varies from aligned to oblique. Oblique whistlers in their turn can very efficiently diffuse suprathermal electrons \citep{Parail78,Vasko_2019}
 Recently, \cite{J.-S.-Halekas:2020aa} analysed the heat flux properties observed by Parker Solar Probe near the Sun (0.125–0.25 AU) and found that its regulation is consistent with oblique whistlers and magnetosonic wave modes.  \\
  \\

Whistler waves therefore are probably an important factor of the solar wind dynamics, although there is no consensus for now about how they are generated under different solar wind conditions and how efficiently they affect the energetic electron population. Solar Orbiter provides us with a great opportunity to explore the role of the whistlers in the solar wind dynamics. \cite{chust2021} (this issue) have analysed in details three wave events observed by Solar Orbiter and found them to correspond to outward whistler waves. In this paper, we extend these results by presenting an overview of the waves observed above 3 Hz by the Solar Orbiter RPW experiment between 0.5 AU and 1 AU during its first orbit.  We characterize the waves in detail and demonstrate that the wind is populated by quasi parallel outward propagating whistler waves. Then we investigate how the whistler waves properties vary with the heliocentric distance. \\

Whistler waves at similar distances were observed by the HELIOS spacecraft in the 70's but with less details. They were analysed first by \cite{1981JGR....86.7755B} and more recently by \cite{2020ApJ...897..118J}. The authors used the data obtained by only two components of the search coil magnetometer, which limited their analysis of the dependence of amplitudes upon different parameters.
Some of our results  show some disagreement with these previous studies. It is worth recalling that there were other statistical studies of whistler waves in the solar wind at  1AU, in particular using CLUSTER \citep{0004-637X-796-1-5} and ARTEMIS \citep{2019ApJ...878...41T} spacecraft measurements. To the exception of \cite{0004-637X-796-1-5} at 1AU, none of these statistical analysis could make a complete polarisation analysis to experimentally demonstrate that the observed waves actually correspond to whistlers waves. We will compare our results with these previous studies when possible.   \\

The paper is organized as follows: sect.\ref{sec_ana} presents the data and the wave detection method; in sect.\ref{sec_pola}, we determine the wave polarisation and propagation direction, discussing the method and instrumental issues in details; sect.\ref{sec_res} presents the variations of the wave properties with the heliocentric distance, and we conclude in sec.\ref{sec_conclu}

  

\section{Data and analysis\label{sec_ana}}

 \begin{figure}
   \centering
    \includegraphics[trim={1cm 3cm 0cm 2cm},width=0.49\textwidth]{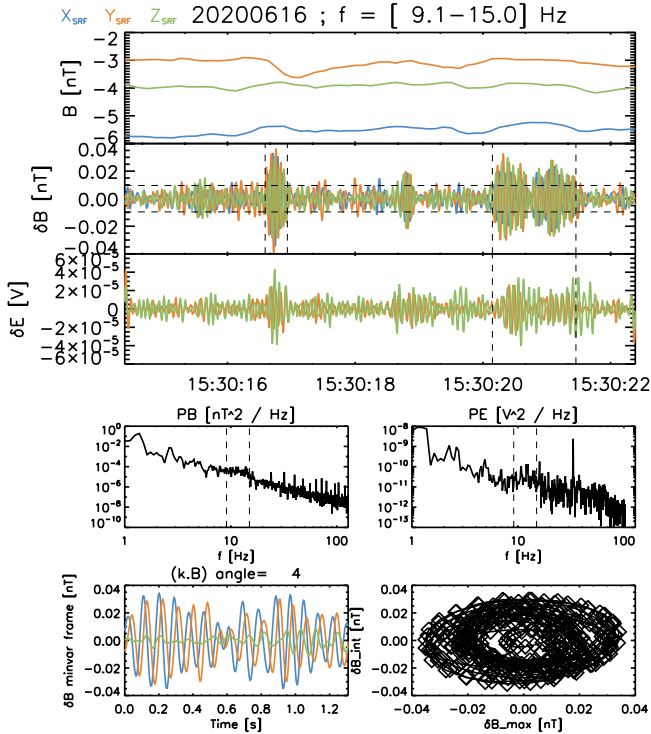}
   \caption{Example of RPW/LFR snapshot at 256Hz and wave packet detection. Three first rows show  the background magnetic field, and the AC magnetic (X, Y and Z components) and electric field (Y and Z components). Horizontal dashed lines shows 3 times the noise level and vertical dashed lines indicate the detected wave packets. Fourth row shows the magnetic and electric power spectra. Fifth row show the minimum variance analysis and hodogram for the longest wavepacket, indicated in the AC electric field panel (third row)  }
              \label{FigSnapshot}%
    \end{figure}

 Our analysis is based on measurements carried out between March 1st, 2020 and December 3rd, 2020 by the magnetic and electric antennas (respectively SCM -Search Coil Magnetometer- \citep{Jannet21} - and ANT ) of the Radio and Plasma Waves (RPW, \cite{2020A&A...642A..12M}) experiment onboard Solar Orbiter. The data covers slightly more than the first orbit covering  distances between 0.51 and 0.98AU from the Sun.
Snapshot waveforms of the fluctuating magnetic field and electric antenna voltage differences are regularly recorded at 256 Hz, 4 kHz and 25 kHz  by the Low Frequency Receiver (LFR). The three electric antenna being in the same plane, we can only access components of the electric field in that plane. Therefore, we will use in this paper the spacecraft reference frame (SRF), that is most often closely related to the RTN frame (X$\sim$-R, Y$\sim$-T, Z$\sim$N). As explained in \cite{2020A&A...642A..12M}, computing the  E$_z$ component requires the combination of two different measurements which makes this quantity less reliable at this early stage of the mission. We therefore focus on the E$_y$ component that is determined as followed : $E_y=-V_{23}/L_y$, where $L_y$ is the effective length which will be discussed later.
Few solar wind signals were recorded by SCM above 128 Hz and we  considered only the snapshot acquired with a sampling frequency of 256 Hz, each of them having a duration of 8 s.\\

At this early stage of the mission, there are two uncertainties in the RPW measurements that we had to deal with: the first one was already pointed out and discussed in details by \cite{chust2021} and concerns a systematic, non frequency-dependent, phase shift (also called phase deviation in the following) between the magnetic and electric measurements. However, because for electromagnetic waves aligned with the background magnetic field, as the ones that we are going to describe, $\delta E$ and $\delta B$ must be perpendicular, this observed phase shift can be identified and corrected, although its origin is still unclear for now. The second uncertainty is the effective length $L_y$ of the electric antenna, which appears to be a difficult quantity to characterize in the interplanetary medium, as reported by \cite{2020JA027980} on  Parker Solar Probe. \cite{steinvall2021} have estimated it for Solar Orbiter by performing a deHoffmann-Teller analysis and comparing velocity measured by the Proton-Alfa Sensor (PAS) of the Solar Wind Analyser \citep[SWA]{SolO_SWA}. We will find similar values of $L_y$ by comparing the theoretical and observed phase velocity of whistler waves in sec.\ref{sec_efflen}, which will allows us to identify the anti-sunward propagation of these waves. \\

We also used the following data, when available,:
\begin{itemize}
    \item the electron density from RPW \citep{Yuri2021}
    \item the measurements of the Solar Orbiter DC magnetometer MAG \citep{SolO_MAG}. This allows us to retrieve the direction of the background magnetic field
    \item the measurements of the solar wind moments acquired by the Solar Wind Analyser (SWA) instrument \citep{SolO_SWA}. 
\end{itemize}  
\subsection{Selection criteria and procedures}
Our objective is to automatically detect the waves and compute their parameters. We have chosen a conservative approach that allowed false detections that we  have removed after-while by applying different additional criteria. We analyzed a snapshot if its 8s average spectrum was at least 3 times larger than the background spectrum in a frequency band of at least 1 Hz. The background spectrum was determined as the daily median spectrum, which gave good results as intervals containing waves occupy only a small fraction of the observations. A similar level of detection was obtained by using a threshold on the coherence between the magnetic components. 
For each 8 s - snapshot with waves, we performed a standard analysis by computing the average magnetic spectral matrix (df=2Hz) and determining the frequency band and maximum frequency of the waves, the $\mathbf{k}$ vector direction (with a $\pm \pi$ ambiguity but $\mathbf{k}$ was initially forced to be in the same half plane as the background magnetic field $\mathbf{B_0}$). We also evaluated the degree of polarisation and the ellipticity of the waves. This was done in two ways, by computing them directly from the observed spectral matrices, as  suggested by \cite{Means1972}, and by doing a singular value decomposition analysis of this same matrix as proposed by \cite{Santolik2003}. The results from the two methods agreed very well with each other and the choice of one or another does not make differences for the statistical results presented here. For each snapshot, we kept the values of the wave parameters obtained at the frequency corresponding to maximum of the normalized spectrum. \\

To gain deeper insight on the observed waves, we added a procedure to detect individual wave packets within a snapshot. We first pass-band filtered the magnetic and electric waveforms in the frequency band determined by analyzing the 8s averaged spectrum. Next, we selected periods where the magnetic field root mean square (rms) was found to be three times higher than the noise level for at least 4 periods, which defines a wave packet. Wave packets separated by less than 1.5 periods were merged. Fig.\ref{FigSnapshot} shows as an example a snapshot observed on June 16, 2020, near 15:30. Two wave packets were detected on that snapshot, while the shorter wave packet around 15:30:19 does not stand enough time above our threshold to be detected. The properties of individual wave packet were determined both by a Fourier analysis to retrieve the wave amplitudes and phases and by a minimum variance analysis  \citep[MVA,][]{Sonnerup1998} of the magnetic waveforms to retrieve the direction of the wave vector  ($\pm \pi$). We also estimated their planarity and ellipticity by using the ratio of the singular values \citep{Santolik2003}.
In comparison to the 8-s average spectral matrix, we can perform a single (non averaged) Fourier transform as the analysis is limited to the time where the wave signal is significant and the incoherent noise to be averaged is therefore less relevant.  \\
The quality of the wave parameters deduced from individual wave packet depends on their duration and signal to noise ratio, meaning that we observed greater dispersion of the value for shorter wave packets. One advantage of analysing individual wave packets resides in the possibility to better determine the temporal filling factor of the waves; we also aimed at performing a more precise analysis of
 faint wave packets whose signal in the spectrum would be relatively weak. \\

The detection method described above sometimes commit errors that may be caused by some spacecraft or instrumental interferences.  We used additional criteria to remove some obvious erroneous detections, caused by signals from the platform or instrumental problems, as well as "waves" with a planarity or ellipticity (as deduced from the average spectral matrices) less than 0.6, or with a phase shift between the maximum and intermediate components in the wave frame different from $\lvert 90 \rvert ^\circ$ by more than 10$^\circ$. By doing so, we focused on circularly polarized waves but the inspection of the events with low value of planarity or ellipticity points preferably towards instrumental interferences rather than actual solar wind signals, although this later explanation can not be ruled out at this stage. Due to the presence of interferences and the low sensibility of SCM at low frequencies, we considered only frequencies above 3 Hz, and finally we kept only the events for which we dispose measurements of the DC magnetic field. 
  \\
  
We ended with 5035 8s-averaged spectra and 17362 associated wave packets, most of which are right handed and quasi aligned with the magnetic field as will be shown in the next section.

\begin{figure}
   \centering
     \includegraphics[width=0.49\textwidth]{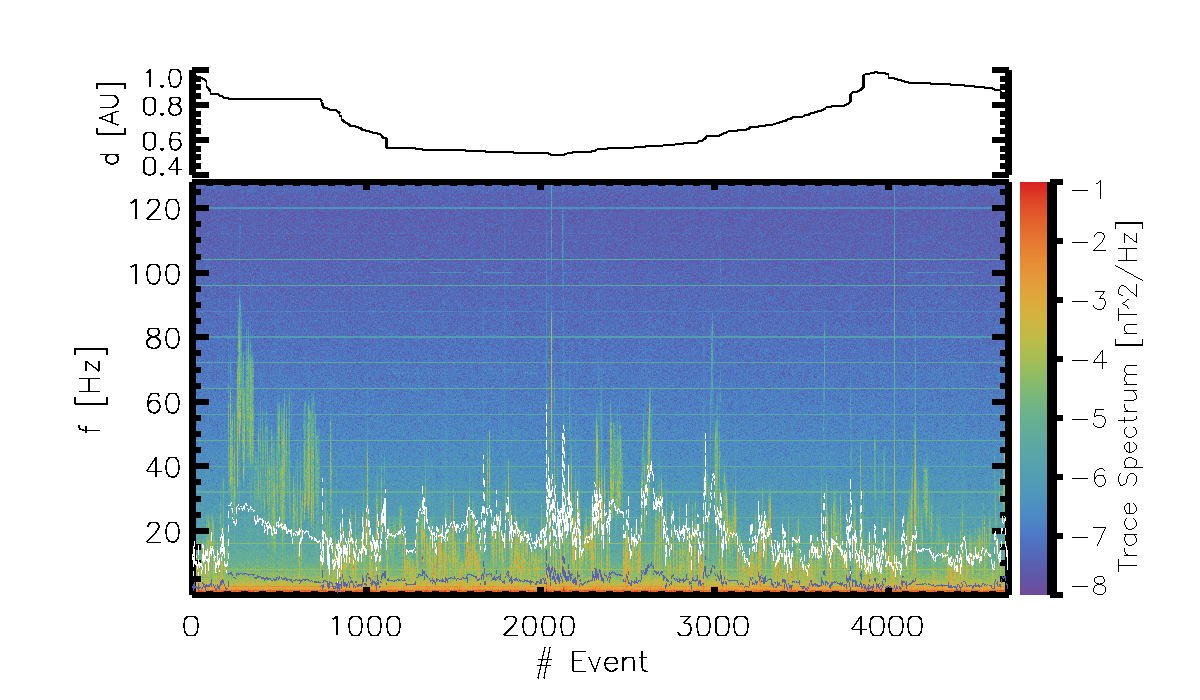} 
   \centering
     \includegraphics[trim={1cm 1cm 0cm 2cm},width=0.45\textwidth]{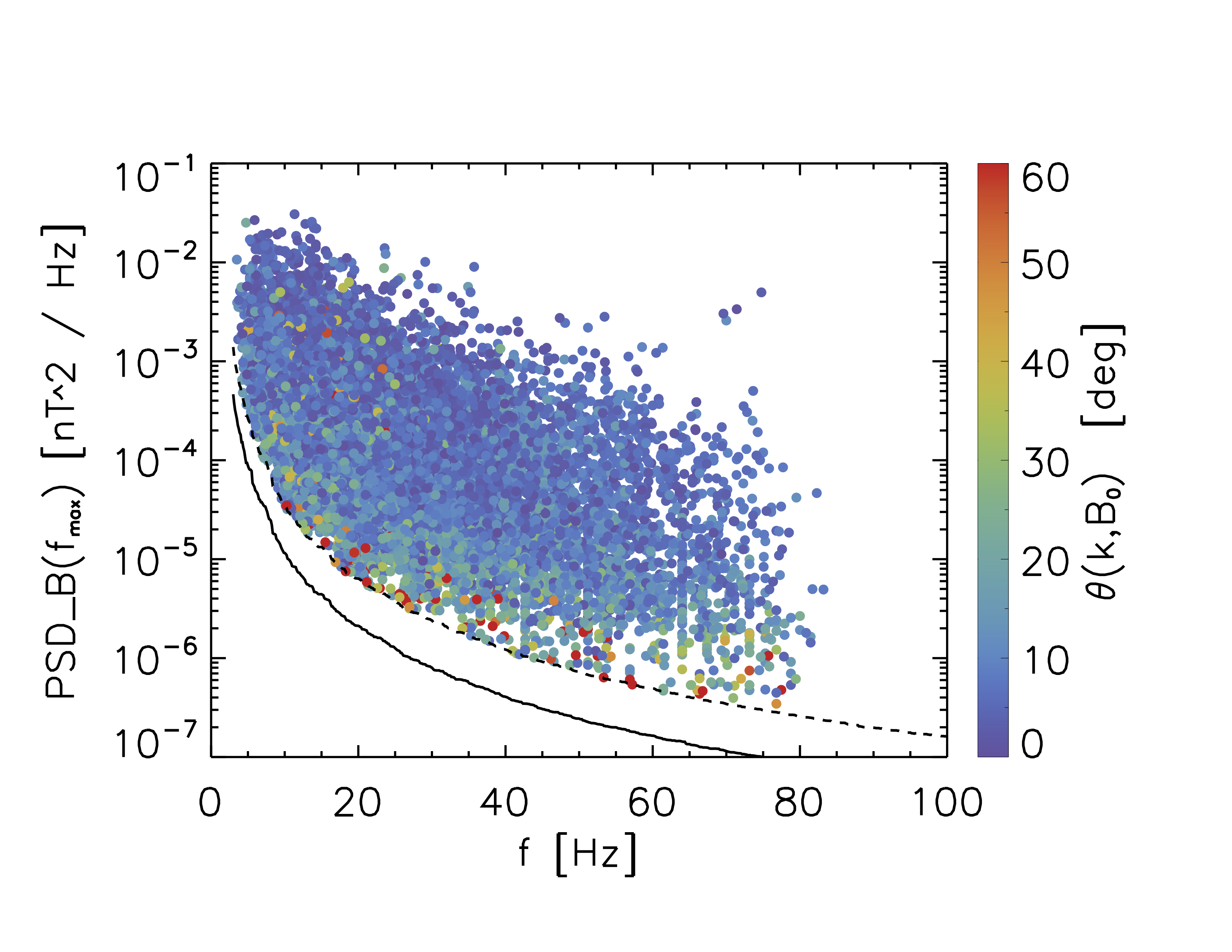} \includegraphics[trim={1cm 1cm 0cm 2cm},width=0.45\textwidth]{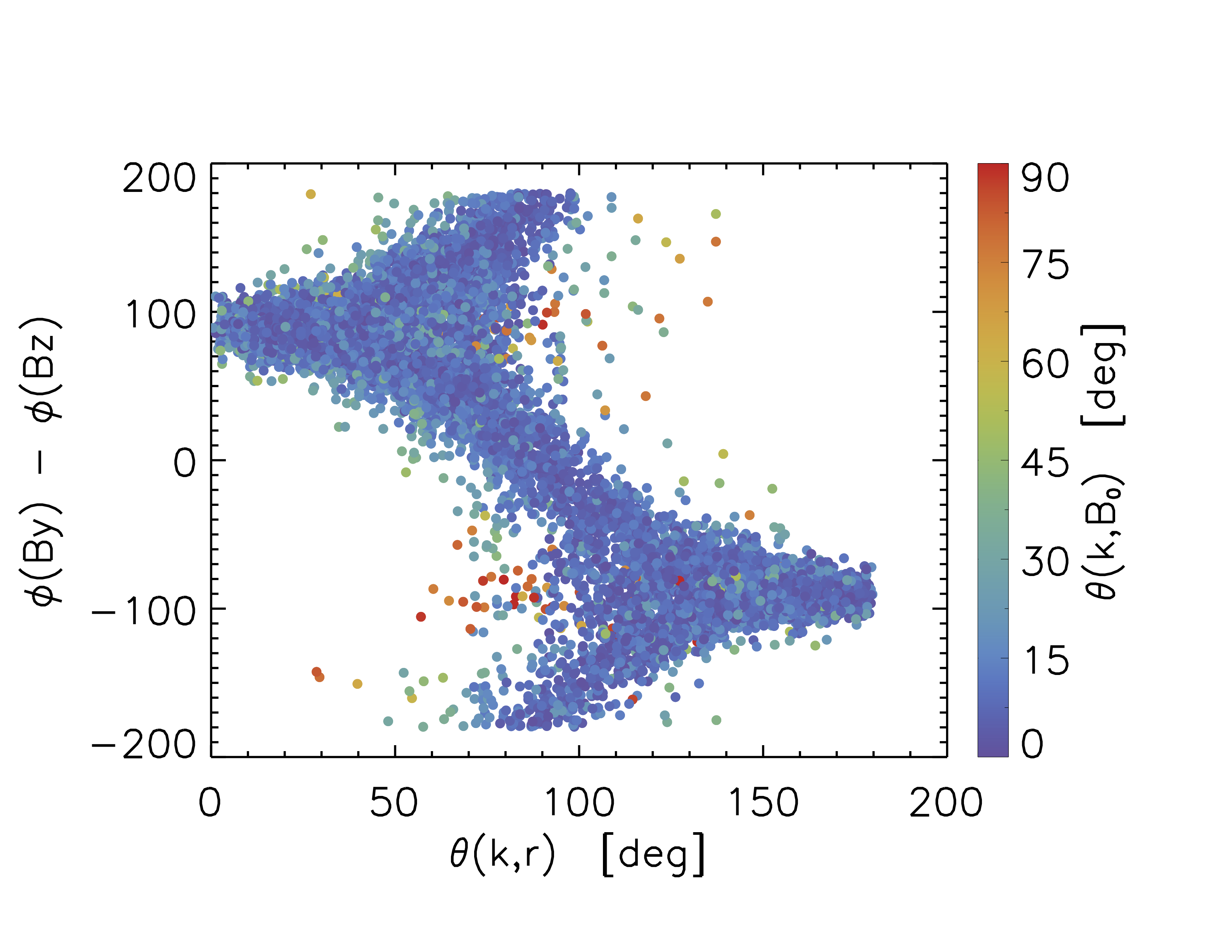}
   \caption{Overview of detected waves. \textbf{Top}: 8 s average trace spectrum for the detected events vs event number, with their distance from the Sun on the top. The dashed white line indicate 0.1f$_{ce}$ and the violet one the lower hybrid frequency f$_{LH}$. \textbf{middle}: Magnetic field PSD at the frequency corresponding to the maximal power for each detected wave packet. The noise level is indicated by the plain black line, and three times the noise by the dashed line. The color indicates the angle between the $\mathbf{k}$ angle and the background magnetic field. 
   \textbf{Bottom}:Phase shift between the $B_y$ and $B_z$ magnetic components versus the angle between the $\mathbf{k}$ vector and the radial direction.  The color indicates the angle between the $\mathbf{k}$ angle and the background magnetic field.}
              \label{FigPhase}%
    \end{figure}

\section{Wave polarisation and propagation direction}\label{sec_pola}
In this section, we analyzed in details the wave properties first in the spacecraft frame and then in the plasma frame in order to identify the wave modes as unambiguously as possible. The top panel of fig.$\ref{FigPhase}$ shows the spectrogram of the detected events, in chronological order but with a non linear time scale; the distance of the Sun at which they were observed is indicated as well. One can note the traces of interferences and artefact signals as horizontal straight lines. Almost no waves were observed in the frequency range above approximately 90Hz and an overhelming majority lies in the frequency range from 0.04 to 0.3 f$_{ce}$ with a maximum around 0.1 f$_{ce}$ (f$_{ce}$ is the electron gyrofrequency) in the spacecraft reference frame. The group of waves with frequencies clearly above 0.1 f$_{ce}$ that occurred at the beginning of the orbit around 0.85 AU corresponds to a strong magnetic perturbation (possibly a magnetic cloud or ICME) that crossed Solar Orbiter on April 13, 2020. We will often treat the waves registered during this perturbation separately, and focus on waves observed during less perturbed time intervals.\\
The middle panel shows the power spectral density of the magnetic field for each wave packet. As can be seen from the color code, most of the waves are quasi aligned with the background magnetic field. A few of them have a significant deviations from parallel propagation: 1.5\% have an angle greater than 40 $^\circ$ and 0.9\% have an angle greater than 60 $^\circ$. It is noticeable that these waves also have a lower amplitude of the magnetic field, as expected for oblique whistler mode waves.  \\

The polarisation of the wave packets in the spacecraft frame is shown in the bottom panel of fig.\ref{FigPhase}. As we noted already, most of the waves are aligned with the background magnetic field, therefore $\theta_{\mathbf{k},\mathbf{r}}\sim \theta_{\mathbf{B_0},\mathbf{r}}$ and the abscissa also indicates approximately the angle between the background magnetic field and the radial direction. 
When the background magnetic field $\mathbf{B_0}$ points in the  direction towards the Sun (small angles), the phase difference between $B_y$ and $B_z$ is positive and close to 90 $^\circ$ while when $B_0$ points outward, the phase  difference is negative. The overwhelming majority of these waves are therefore right handed (RH) in the spacecraft frame. 
Evaluation of the phase difference between the largest and middle eigenvectors in the MVA frame (not shown here), with the k-vector and background magnetic field vector assumed to be in the same half space with respect to the plane formed by these eigenvectors, confirms that the waves are right handed ($\sim +90^\circ \pm 10 ^\circ$). Only 0.6\% of the events (109 packets) have left handed (LH) polarisation in the spacecraft frame, but these are also the waves that were found to be more oblique: 73\% of them have a ($\mathbf{k}$,$\mathbf{B_0}$) angle above 50$^\circ$. By contrast, only 0.7\% of the RH polarized waves have a ($\mathbf{k}$,$\mathbf{B_0}$) angle above 50$^\circ$. We leave these interesting but non-typical events for future studies and hereafter concentrate on relatively field aligned  ($\theta_{\mathbf{k},\mathbf{B_0}}<40$ $^\circ$) and right handed polarized waves, which are good and sound candidates for being whistler waves. \\

Their observed frequencies and RH polarisation correspond indeed to the frequency range and polarisation of whistler waves. However, these observed properties are Doppler shifted by the solar wind velocity and the actual identification of the waves requires to determine them in the plasma reference frame. As we encountered two instrumental difficulties, which is not surprising at this stage of the mission, the phase deviation between E and B and the value of the effective length of the electric antenna, we think important to describe how we tackled them in details. We first give a brief outline of the method before describing it in details in the next sections :
\begin{itemize}
    \item we inspected the phase difference between $E_{y}$ and $B_{z}$ to investigate whether the quasi parallel waves propagates sunward (equivalently inward) or anti-sunward (equivalently outward). This leads us with two possible scenario for the phase deviation corresponding to the two propagation directions, which will be resolved by comparing the effective lengths.
    \item for both scenario, we determined the wave frequency and theoretical phase velocity in the plasma frame by applying the Doppler shift and using the plasma dispersion relation for whistler mode waves.
    \item we compared the observed and theoretical wave phase velocities to determine the effective length in both scenario.
    \item we found the anti-sunward scenario to be the only one that provides realistic effective lengths in excellent agreement with an independent analysis using the deHoffman-Teller frame.  
\end{itemize}
By anticipation, we note that our correction of the E and B phase deviation and our derived relation for the effective length antenna, which leads us to the conclusions that the observed waves are anti-sunward propagating whistler waves, are respectively consistent with the studies of \cite{chust2021} and \cite{steinvall2021} and appear therefore reliable.

\subsection{Doppler shift}  
Determining  the wave properties in the plasma reference frame requires to perform the Lorentz transformation of the fields, which results from the wave being convected by the solar wind as seen from the spacecraft.  If the waves propagate outward, the frequency in the plasma reference frame increases and the polarisation remains unchanged with respect to the observations in the spacecraft reference frame. On the other hand, if the waves propagate inward, the frequency decreases and the polarisation may also change if the phase velocity of the wave is smaller than the solar wind speed. Taking into account the observed frequency and  polarisation in the spacecraft frame, the observed RH waves can be 
\begin{itemize}
    \item outward propagating whistler waves
    \item inward propagating whistler waves with $v_{\varphi}>v_{sw}$. The waves are therefore inward propagating in the spacecraft frame.
    \item inward propagating ion cyclotron waves with $v_{\varphi} <v_{sw}$. The waves are therefore outward propagating in the spacecraft frame 
\end{itemize}
Here, $v_{\varphi}$ stands for the phase velocity of wave.
The phase speed of ion cyclotron waves (ICW) is close to the Alfv\'en velocity $V_A$. In our sample, assuming $n_p=n_e$ and using the electron density from the spacecraft potential, $V_A$ has a median value of 36 km/s and a standard deviation of 17 km/s. It is therefore always well below the solar wind speed, and inward ICW could be observed with a RH polarisation in the spacecraft frame with an outward velocity $v_{sw}-V_A \sim 0.9v_{sw}$ close to the solar wind speed. 
Inward whistler mode waves should be observed with an inward velocity $v_\varphi-v_{sw}$ and outward whistler mode waves with an outward velocity $v_\varphi+v_{sw}$. \\

Therefore, even with an ambiguity in the determination of the wave propagation direction caused by the uncertain phase difference between E and B (the "phase deviation" problem), the analysis of the absolute phase speed velocity and its comparison to theoretical expectations can provide us with a strong argument in favor of one of the two options proposed above. 

\subsection{Wave propagation direction}
  \begin{figure}
   \centering
    \includegraphics[trim={0cm 1cm 0cm 2cm},width=0.49\textwidth]{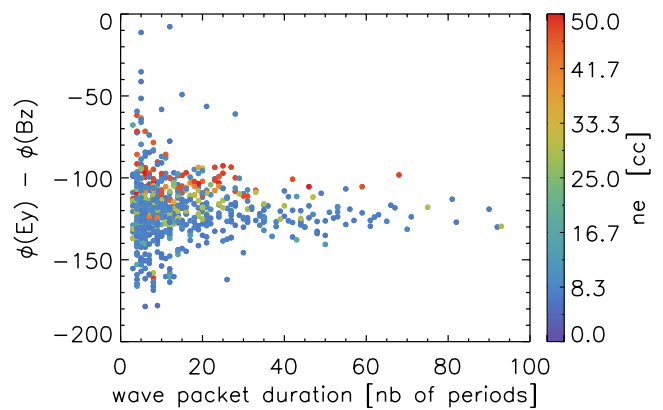}
    \caption{Phase difference between E$_y$ and B$_z$, \textit{vs} wave packet duration and electron density deduced from the spacecraft potential.}
              \label{FigEBPhase}%
    \end{figure}
    
The wave propagation in the plasma frame is the direction of the Poynting flux; it can be determined by using conjointly the magnetic and electric fields measurements. RPW can only determine the Y and Z  components of the electric fields E$_y$ and E$_z$, which are in a plane perpendicular to the X direction which is most often the radial direction. At this stage of the mission, the Y component is the most reliable as it is determined directly from the voltage difference between two symmetric antennas, the evaluation of the Z component requiring  a combination of  measurements coming from the three antennas. As we are dealing with plane waves aligned with the background magnetic field, we stress that only one component of the electric field is necessary to remove the $\pm \pi$ ambiguity on the waves propagation direction determined with the magnetic field.  As indicated earlier, $E_y=-V_{23}/L_y$. The value of $L_y$ can depend on plasma conditions and is known to be difficult to assess; it affects the absolute value of the electric field estimate but not its phase. We first discuss the propagation direction of the wave.\\

 The projection of the Poynting vector of the waves in the X direction depends on the product $E_y.B_z$ and therefore on the phase difference between $E_y$ and $B_z$. For a plane wave parallel to the background magnetic field that is itself aligned with the radial direction ($\pm$ X), the phase difference between  $E_y$ and $B_z$ must be either 0 $^\circ$, if the wave propagates towards the Sun ($\vec{E_y} \times \vec{B_z}$ in the +X direction), or 180 $^\circ$ if the wave propagates outwards. Fig. \ref{FigEBPhase} shows the phase difference $\varphi_{E_y} - \varphi_{B_z}$ for right handed circular polarized waves that propagates parallel (or anti parallel) both to the background magnetic field $\mathrm{B_0}$ ($\theta_{\mathbf{k},\mathbf{\pm B_0}}< 20 ^\circ$) and to the radial direction ($\theta_{\mathbf{k},\mathbf{\pm X}}< 20 ^\circ$). 
 It is plotted as a function of the wave packet duration, that we use as a criterion for signal to noise ratio. The color indicates the electron density. One can see that the phase difference is nearly constant and approximately equal to -130 $^\circ$; it does not appear to vary with the frequency (not shown), indicating a propagation in the half plane anti sunward to the Sun. The first conclusion is therefore that all these waves propagate in the same direction. We also note a tendency for the phase  difference to slightly increase for larger densities, the reason for this being unclear at this point. However, this value of -130 $^\circ$ does not match the wave vector direction found with the magnetic field, for which one should find either -180 $^\circ$ or 0 $^\circ$. Our interpretation of this unexpected phase deviation, similar to that deduced by \cite{chust2021} based on the detailed analysis of a few single whistler events, is that this is caused by an unsolved instrumental problem. 
 Since the phase deviation is well established for the parallel (or anti parallel) whistler waves, our conclusion is that there is a constant error in the registration of the phase of one of the two signals, which let us with have two possible scenario:
\begin{enumerate}
    \item a "weak phase deviation", for which E$_y$ (\textit{resp., }B$_z$) should be delayed (\textit{resp., } advanced) by removing (\textit{resp., } adding) 50 $^\circ$, so that the phase difference matches 180 $^\circ$. This will lead to the conclusion that the waves propagate outwards. 
    \item a "strong phase deviation", for which E$_y$ (\textit{resp., }B$_z$) should be advanced (\textit{resp., } delayed) by adding (\textit{resp., } removing) 130 $^\circ$, so that the phase difference matches 0 $^\circ$. Such phase difference will correspond to waves that propagate inwards.
\end{enumerate}
The same conclusion can be drawn by analyzing the phase difference between E$_y$ and $B_y$, which has to be equal to $\pm$ 90$^\circ$. We present below some extra argument that indicates that the first option is very likely correct. 
In any case, it is important to notice that the (B$_z$,E$_y$) phase difference remains constant for practically all the parallel waves and frequencies, which provides a strong argument to suggest that the large majority of waves observed between 1AU to 0.5AU are propagating in the same direction. \\



\subsubsection{Wave frequency in the plasma frame}

The observed frequency in the spacecraft frame (superscript $^{SC}$, quantities in the solar wind frame have no superscript) $\omega^{SC}$ has been Doppler shifted by the motion of the solar wind so that:  
  \begin{equation}\omega = \omega^{SC} - \vec{k}.\vec{V_{SW}}
\end{equation}
Or, supposing that both $k$ and $V_{SW}$ are positive quantities,
\begin{equation}
\omega_{in,out} = \omega^{SC} \pm kV_{SW}\lvert \cos{\theta_{\vec{k},\vec{V_{SW}}}}\rvert
\end{equation}
where the subscripts $_{in}$ and $_{out}$ stand respectively for inward (sign +) and outward (sign -) propagation. 
Combining this equation with the following  plasma dispersion relation for whistler mode waves in a  cold plasma approximation,
  \begin{equation}
k=\frac{\omega_p}{c}\sqrt{\frac{\omega}{\Omega_c\cos\theta - \omega}}\label{eqdispersiong}
\end{equation}
one can derive the frequency in the plasma frame 
  \begin{equation}
\omega_{in,out} = \omega^{SC} \pm V_{SW}\lvert \cos{\theta_{\vec{k},\vec{V_{SW}}}}\rvert \frac{\omega_p}{c}\sqrt{\frac{\omega_{in,out}}{\Omega_c\cos\theta - \omega_{in,out}}}
\end{equation}
where $\pm$ corresponds to inward and outward propagation respectively. We solved this expression numerically using the Brent's method \citep{Brent:1973aa}. We used the electron density  determined by RPW from the spacecraft potential measurements, and the solar wind velocity measured by SWA whenever available, which is rare (SWA has unfortunately not been working continuously): in most of the cases, we had to assume a purely radial solar wind speed of 350 km/s. The validity of this assumption will be discussed later.
Once the frequency in the plasma frame determined, one can compute the theoretical wave vector and phase velocity $v_\varphi=\frac{\omega}{k}$ in the two scenario.


\subsubsection{Phase velocity in the plasma frame}
The observed phase velocity $v_{\varphi}=\omega/k$ can be computed by considering the Faraday equation for waves, 
\begin{equation}
\mathbf{k}\times\mathbf{E}=\omega\mathbf{B}
\end{equation}
, which easily leads to the following expression \citep{chust2021}      
\begin{equation}
\hat{v}_{\varphi}=\frac{\mathbf{n}.\mathbf{B_0} (\hat{E_y}\hat{B_y^*})}{B_{0x}(\hat{B_z}\hat{B_y^*}) - B_{0z}(\hat{B_x}\hat{B_y^*})}
\label{eqvphi2g}
\end{equation}
where $\mathbf{n}=\frac{\mathbf{k}}{k}$, $\mathbf{B_0}$ is the background magnetic field, $\hat{}$ denotes the complex amplitude and $^*$ the complex conjugate. \\
The equation is valid for a plane wave when there is no parallel fluctuating electric field, which is our case here. It allows us to consider non purely radial propagation, but gives the same results $\hat{v}_{ph }=\hat{E}_y/\hat{B}_z$ in this latter case. The multiplier $\hat{B_y^*}$ could be removed but it allows us to work with relative phases more easily;  the cross terms $\hat{B_i}\hat{B_j^*}$ or $\hat{E_i}\hat{B_j^*}$ are averaged when computed from the averaged spectral matrix obtained from the 8s snapshot. \\

$v_\varphi$ as expressed in Eq.\ref{eqvphi2g} is in the solar wind frame if one uses the electric field in the solar wind frame. This latter can be determined by considering that the solar wind frame is moving with $v_{sw}$ in the -X$_{SRF}$ direction as seen from the S/C frame and applying the classical Lorentz transformation and :
\begin{equation}
\mathbf{E}=\mathbf{E^{sc}}+\mathbf{v}\times\mathbf{B}
\end{equation}
where $\mathbf{E^{sc}}$ is the electric field measured in the spacecraft frame. For $E_y$, using complex notation for fluctuating field $E_y=\Re(\hat{E_y}e^{ikr-i\omega t})$, this leads to
\begin{equation}
\hat{E}_y=\hat{E}^{sc}_y - v_x \hat{B}_z = \hat{E}^{sc}_y+v_{sw}\hat{B}_z
 \label{EqConvEg}    
\end{equation}
where we have assumed that the wind flows in the radial direction only.\\
Equivalently, one can compute the phase velocity in the spacecraft frame by using eq.\ref{eqvphi2g} with $E_y=E^{sc}_y$ and then remove $\vec{k}.\vec{V_{SW}}/k$ from the result. \\
Let us emphasize that the observed wave phase velocity is anti correlated with the effective length $\hat{E}^{sc}_y=\frac{\hat{V}^{sc}_y }{L_{eff}}$, where $\hat{V}^{sc}_y$ is the fluctuating electric potential observed in the spacecraft frame in the Y direction.

\subsection{Effective length of electric antenna}\label{sec_efflen}

 \begin{figure}
   \centering
    \includegraphics[trim={0cm 1cm 0cm 2cm},width=0.49\textwidth]{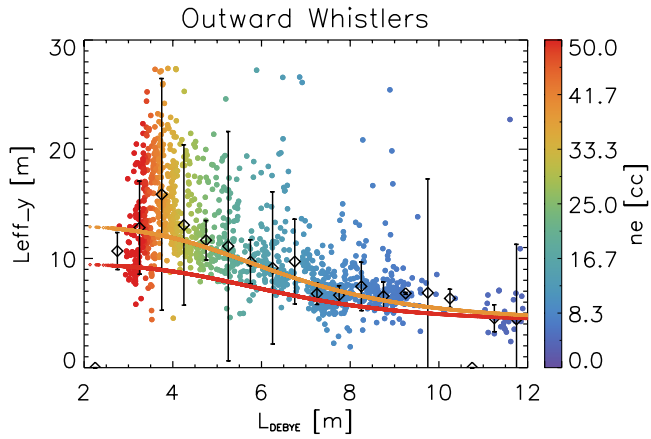} 
    \caption{Effective length considering outward propagating whistlers. The relation obtained by Steinvall et al. (this issue) using deHoffmann-Teller analysis is in red, the modified one ($L_{eff},{max}$=13 m) in orange.} 
              \label{FigLeff}%
    \end{figure}

At this stage of the mission,  the effective length of the antenna is not yet unambiguously known. Its value is important as it affects the evaluation of the amplitude of the electric field and therefore the estimate of the wave velocity.  We can estimate the effective length by comparing the measured phase velocity in the solar wind frame and the expected velocities obtained from the dispersion relation (Eq.\ref{eqdispersiong}). The measured velocity is obtained by using Eq.\ref{eqvphi2g} with the electric field in the solar wind frame (Eq.\ref{EqConvEg}) and by taking into account the corrections discussed above for the phase deviation
The relation $v_\varphi=\frac{\omega}{k}$ leads to the following equation for the effective length estimate
  \begin{equation}
L_{eff}=\lvert  \frac{\hat{V}^{sc}_y\hat{B}^*_y}{\frac{\omega}{k}\frac{B_{0x}(\hat{B_z}\hat{B_y^*}) - B_{0z}(\hat{B_x}\hat{B_y^*})}{\mathbf{n}.\mathbf{B_0}}-V_{sw}\hat{B_z}\hat{B_y^*}}\rvert
  \end{equation}
This equation can be separately applied to inward ($\varphi_{corr}=130 ^\circ$ and $\frac{\omega}{k}=\frac{\omega_{in}}{k_{in}}$) and outward ($\varphi_{corr}=-50 ^\circ$ and $\frac{\omega}{k}=\frac{\omega_{out}}{k_{out}}$) propagating whistler mode waves. The measured effective lengths, derived for outward propagating whistlers propagating at less than 60$^\circ$ from the radial direction (to avoid low $E_y$ value) and computed from the 8s averaged spectra, are shown in fig.\ref{FigLeff}. There is a very good agreement with a different and independent estimate of the effective length, represented in red and obtained by \cite{steinvall2021}; this latter was derived by matching the velocity of the deHoffmann-Teller (HT) frame, deduced from RPW measurements of the DC electric and magnetic fields, to the measurements of the solar wind velocity by SWA/PAS. The curve in orange is a variation of the Steinvall et al. equation, by taking $L_\mathrm{eff,max}$= 13 m instead of 9.5 m in their equation 3 : $L_\mathrm{eff}=L_\mathrm{eff,min}+\frac{L_\mathrm{eff,max} - L_\mathrm{eff,min}}{1+(\lambda_d/L_\mathrm{antenna})^4} $, where  $\lambda_d$ is the Debye length. Assuming inward propagating whistler waves leads to much longer nonphysical effective lengths, in clear disagreement with the deHoffmann-Teller analysis. This can be understood since in this case the phase velocities in the spacecraft frame would be significantly smaller (only inward whistlers with phase velocities faster than the wind speed keep the RH polarisation in the spacecraft frame), which would therefore require a larger $L_{eff}$ (smaller $E_y$) to match the expected phase speed $\frac{\omega}{k}$. There is still a significant dispersion of the points in our estimate of the effective length, and additional work and more statistics are required to obtain a definitive value that can be used as a standard for the calibration of the electric field ; this will be done in future studies.  \\

Nevertheless, the very good agreement between the evaluations by two independent methods  of the effective length of the antenna provides additional argument in favor of the weak phase deviation corresponding to outward propagating whistler waves. Furthermore, since we can now better cross calibrate the electric and magnetic field by applying the phase correction (-50$^\circ$), and have a reliable estimate of the effective length, one can get a more  reliable estimate of the phase speed.

\subsection{Wave velocity and mode identification}
 \begin{figure}
   \centering
    \includegraphics[trim={0cm 1cm 0cm 2cm},width=0.49\textwidth]{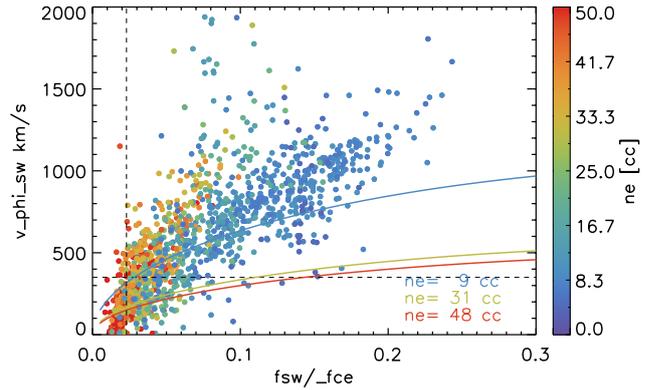} 
    \caption{Phase velocity in the solar wind frame. Theoretical velocities for quasi aligned whistler with $f_{ce}\sim$ 190 Hz and different value of electron density are shown for reference as plain colored lines. The vertical dashed lines indicates the lower hybrid frequency, the horizontal dashed line the typical solar wind velocity}
              \label{Figvphi}%
    \end{figure}

The phase velocities evaluated in the solar wind reference frame, obtained by making use of the electric and magnetic fields measurements and taking into account 
the modified Steinvall et al. relation for the effective length ($L_{eff,max}=$ 13m instead of 9.5m), are presented in fig.\ref{Figvphi} for the same set of the events as presented in fig.\ref{FigLeff}.
The observed values vary in the range from less than 100 km/s to more than 1000 km/s, and can thus be both larger or smaller than the velocity of the solar wind. The velocity obtained using the relation dispersion for three cases are plotted for reference; the agreement with the theoretical expectations obtained by using the observed value of $f_{ce}$ and $f_{pe}$ for each single point, not shown for clarity, is excellent below $n_e \sim$ 30 cm$^{-3}$. \\

\begin{figure*}
   \centering
    \includegraphics[trim={0cm 1cm 0cm 2cm},width=0.99\textwidth]{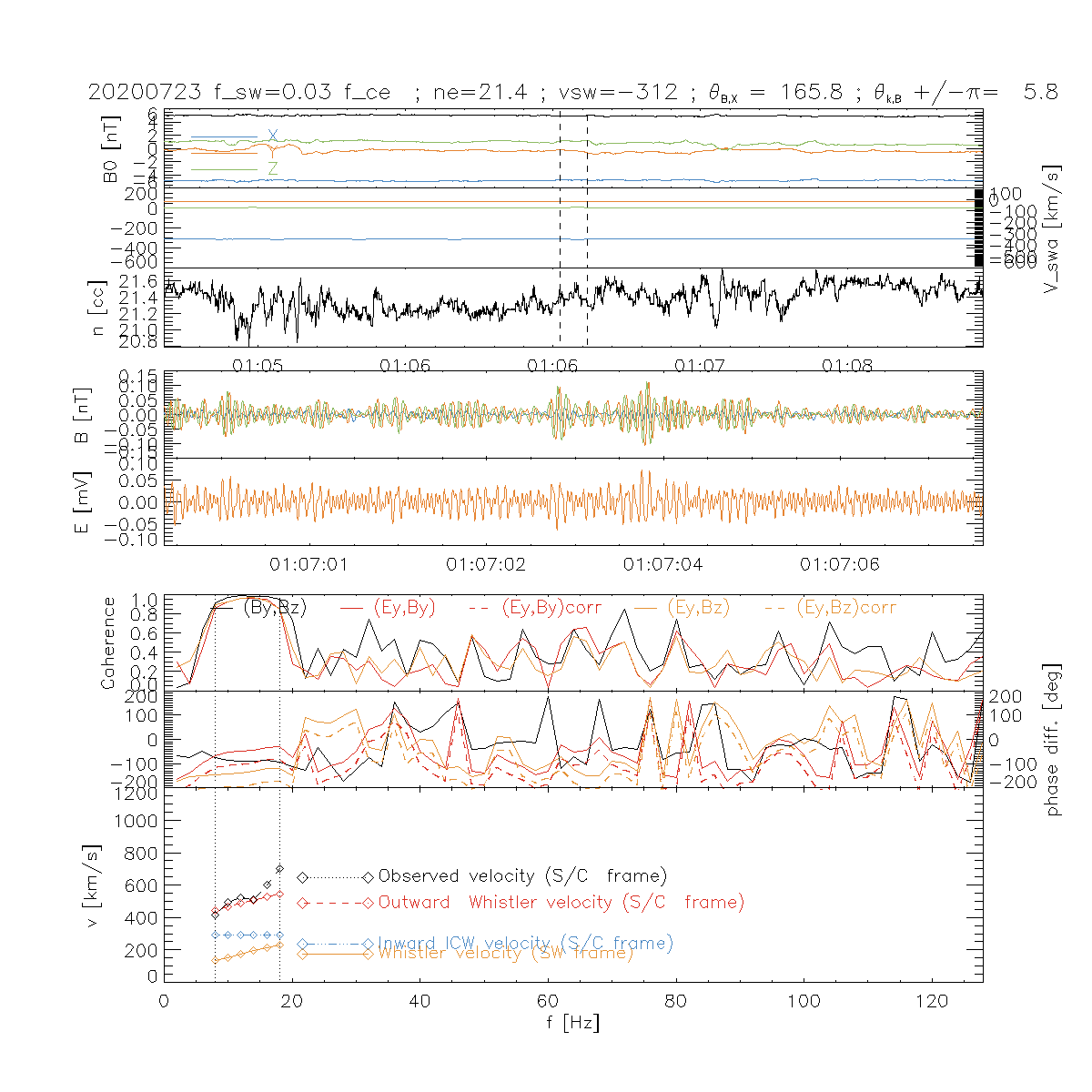} 
    \caption{Wave example 1. The tree first panels starting from top represents de background magnetic field, the solar wind speed, and the electron density measured by RPW, for a 4 minutes interval. centered on the snapshot (the vertical dashed lines indicates the snapshot boundary).The 4th and 5th panel show the 3 components and the Y component of the fluctuating magnetic and electric fields respectively. X is in blue, Y in  orange, and Z in  green. The three bottom  panels show  the spectral coherence, the  phase difference (with electric field corrected by -50$^\circ$ in dashed lines)  and the expected (orange, blue  and  red) and measured (black) the wave phase velocity. The dotted  vertical lines indicates the frequency range where the computation of the velocity is reliable (coherence greater than 0.8 between By and Bz and greater than 0.7 between Ey and Bz)}
              \label{Ex1}%
    \end{figure*}

\begin{figure*}
   \centering
    \includegraphics[trim={0cm 1cm 0cm 2cm},width=0.99\textwidth]{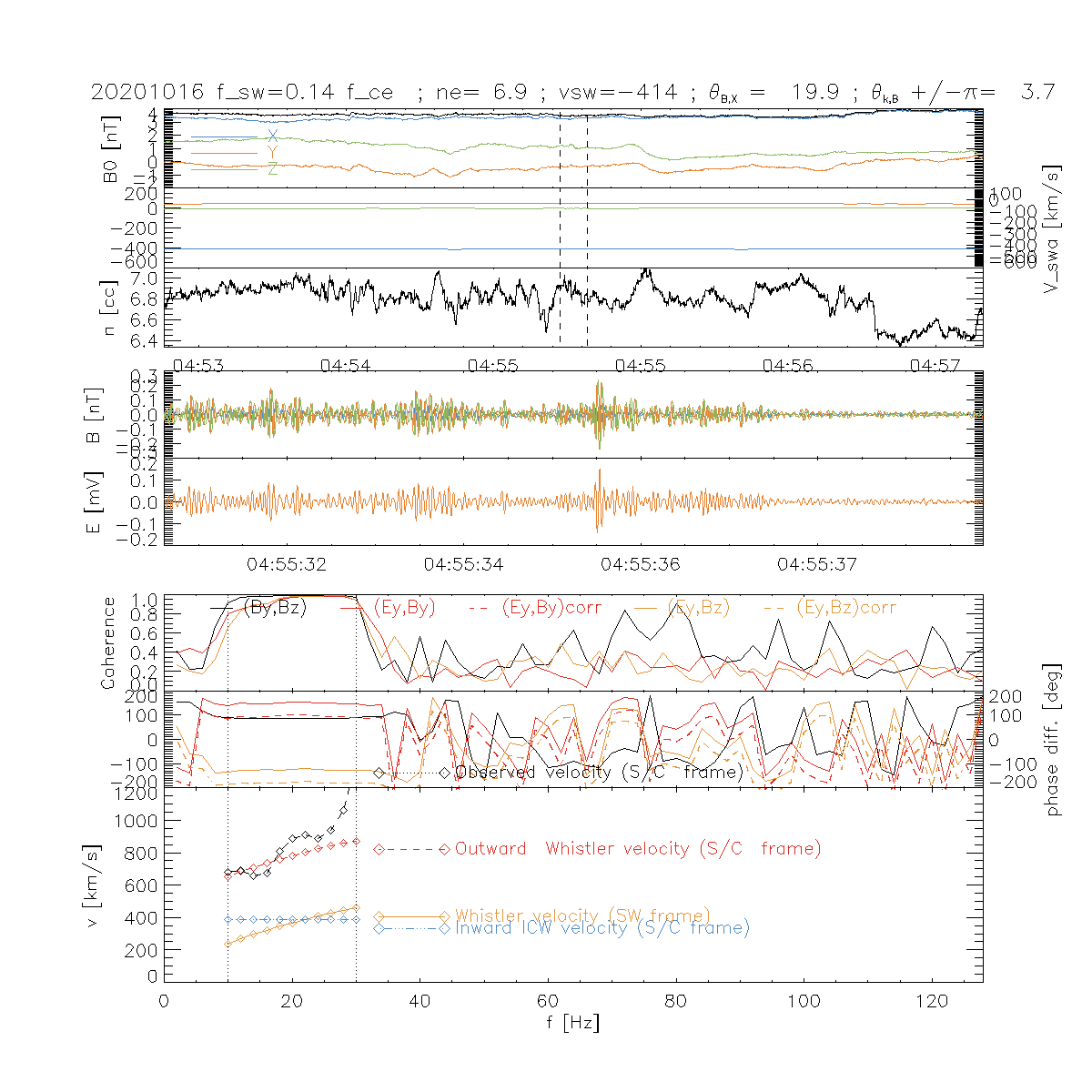} 
    \caption{Same as fig.\ref{Ex1} for wave example 2}
              \label{Ex2}%
    \end{figure*}

\begin{figure*}
   \centering
    \includegraphics[trim={0cm 1cm 0cm 2cm},width=0.99\textwidth]{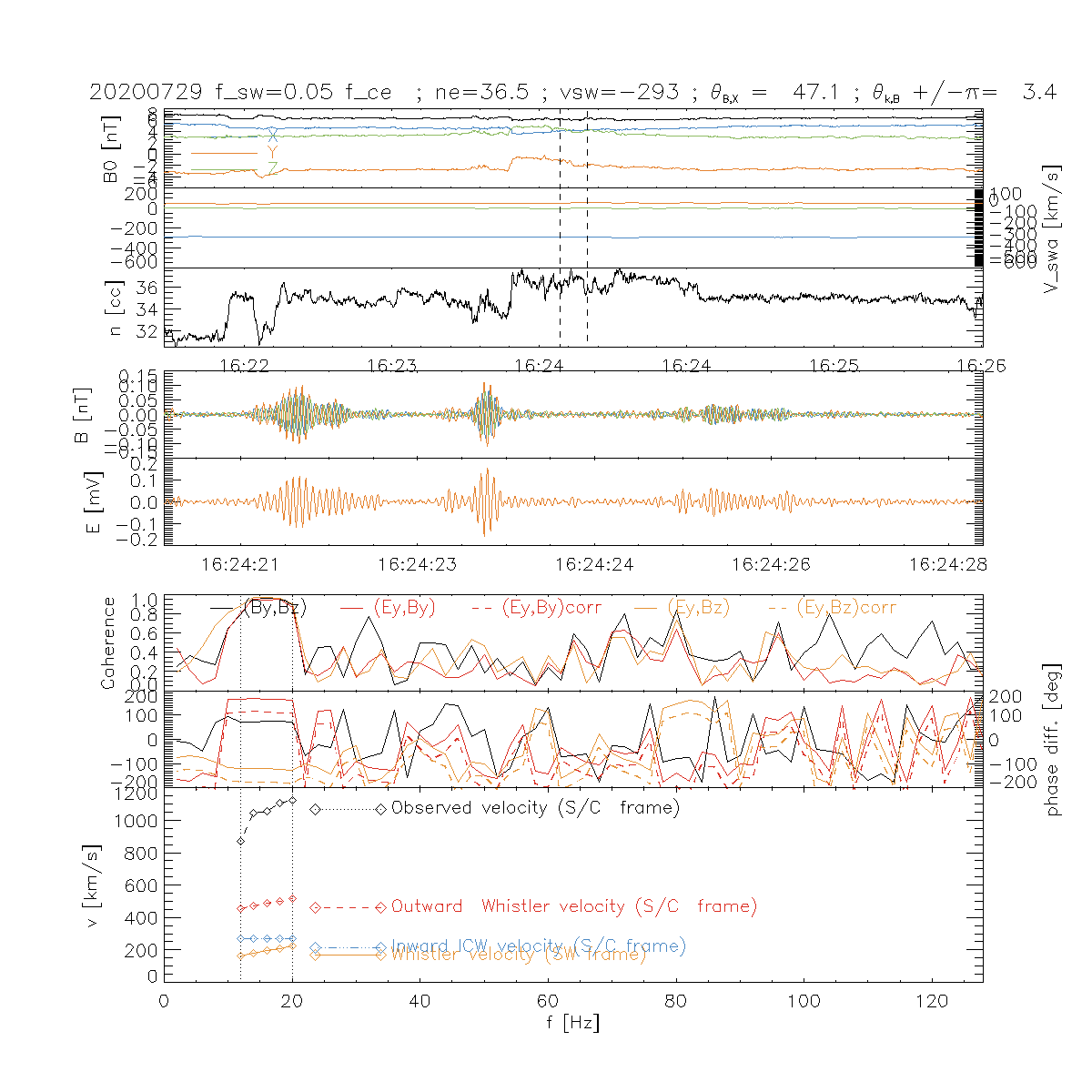} 
    \caption{Same as fig.\ref{Ex1} for wave example 3}
              \label{Ex3}%
    \end{figure*}

\begin{figure*}
   \centering
    \includegraphics[trim={0cm 1cm 0cm 2cm},width=0.99\textwidth]{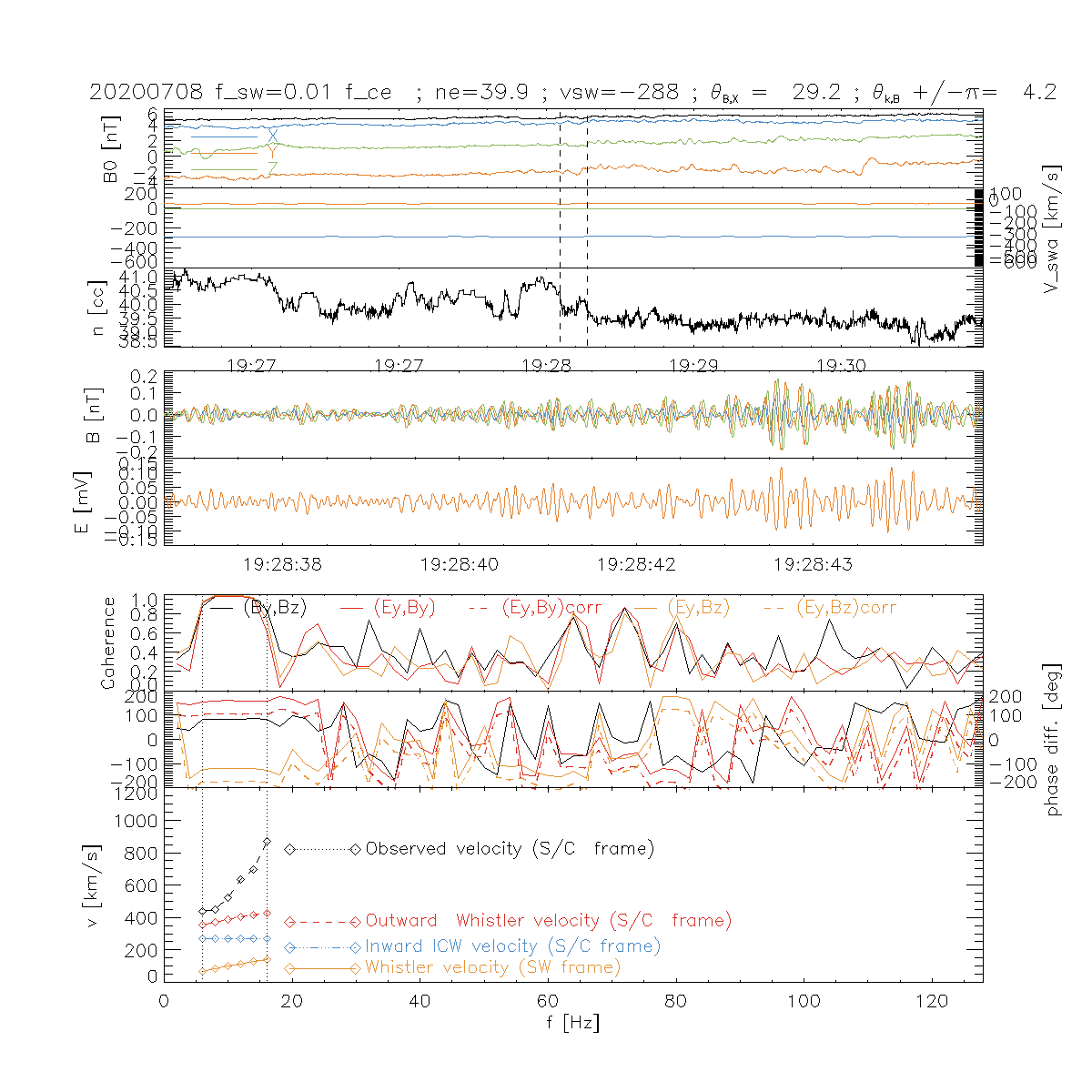}
    \caption{Same as fig.\ref{Ex4} for wave example 4}
              \label{Ex4}%
    \end{figure*}


At density above $\sim$ 30 cm$^{-3}$ corresponding to Debye length of $\sim$ 4m, as could be anticipated from fig.\ref{FigLeff}, the observed velocities are often larger than expected. To investigate whether this compromise the identification of the waves as whistler mode, we show on  fig.\ref{Ex1} to fig.\ref{Ex4} the detailed analysis for four representative cases for which plasma parameters were available. For each figure, the three first panels show the context of the wave observation, the two next panels the magnetic and electric filtered waveforms, and the three last panels the coherence, phase difference, and the observed and theoretical wave phase velocity. The phase difference are plotted without and with (in dashed line) the correction of -50$^\circ$ for the electric field. The observed wave phase velocity is computed using the effective length relation derived in the previous section (orange curve on fig.\ref{FigLeff}). \\
The two first examples in fig.\ref{Ex1} and fig.\ref{Ex2} show cases at low and intermediate density for which the agreement between the expected and observed phase velocity for outward propagating whistler is very good. The correction for the phase deviation allows to retrieve the expected behavior, in agreement with the magnetic field, for an aligned RH wave that propagates anti-sunward. It is clear that the observed right handed circular polarisation of the wave cannot be attributed neither to an inward ion cyclotron wave (in blue) or an inward whistler wave (not shown but with smaller velocity $v\varphi - V_{SW}$), as this would correspond to smaller phase velocity and therefore require longer effective length by a factor 1.5  to 2. The third example on fig.\ref{Ex3} shows a high density case where the  coherence  and phase difference are similar than in the previous example but for which the observed velocity is larger than expected, by a factor $\sim$ 2. It is representative of the points forming the knee of the effective length at $L_{DEBYE}\sim$ 4 m in fig.\ref{FigLeff}. As can be seen, the high velocity is incompatible with an ICW propagating towards the Sun  -and making it compatible would require a very nonphysical effective length-. We don't know other wave modes that whistler to explain these wave properties, and rather tentatively attribute the discrepancy in velocity to other effects: finite plasma beta or some plasma-antenna interactions as the Debye length becomes short.   \\


We also notice that some observed waves (about 7\%) have frequencies below the lower hybrid frequency $f<f_{LH}$. These waves are mostly observed in the regions with high electron density electron densities are higher.
However, there are only 22 cases with $f_{sw}<f_{LH}$ for which the frequency in the plasma frame was computed using the observed solar wind speed and not the default value of 350 km/s. A lower actual velocity would lead to increase the frequency. Fig.\ref{Ex4} show a wave event with high density and at low frequency (0.14 f$_{ce}$). The observed velocity is 1) too high to be caused by an inward ICW propagating in a solar wind speed with a bulk velocity of 288 km/s and 2) increasing with frequency, which is expected for whistler waves but not of ICW. All these waves with frequency below f$_{LH}$ (but one) have velocities larger than the local Alfvén speed in the solar wind frame, which does not favor an interpretation in term of ICW. Here again, we cannot see other wave modes than whistler to explain these observations.  \\
Let us further note that observing low values of $f/f_{ce}$ for whistler waves is not rare. \cite{2020ApJ...897..118J} with Helios data covering about the same range of heliocentric distances as here, found a mean frequency to be around 0.1 $f_{ce}$ in the spacecraft frame, but also a significant number of waves with frequencies smaller than  0.05 $f_{ce}$. Similar distribution of $f/f_{ce}$ in the spacecraft frame were found with Parker Solar Probe data by \cite{2021arXiv210106723J}. \cite{0004-637X-796-1-5} reported examples of whistler waves occurring around 0.03-0.04 f$_{ce}$ as well; \cite{Stansby_2016} have compared observed and theoretical whistler dispersion relation in the solar wind at 1 AU and found whistler waves with frequencies between $\sim$0.03 $f_{ce}$ and 0.2$f_{ce}$. The values obtained here are therefore consistent with these previous results.\\



Previous statistical studies of whistler waves below 1 AU were based on Helios and Parker Solar Probe onboard computed spectra only \citep{2020ApJ...897..118J,2021arXiv210106723J} and were not able to determine the polarization and velocities of the waves, nor the effect of the Doppler shift. We believe that the detailed analysis presented here strongly suggest for the first time that the solar wind between 0.5 and 1AU is populated with outward and quasi parallel propagating whistler waves.

\section{Dependence of the whistler statistical properties properties on the heliocentric distance }\label{sec_res}
We now investigate how the whistler wave parameters vary with the distance. Ideally, we should also take into account the variability of different solar wind parameters, in particular the solar wind velocity (slow or fast), but also the electron beta and temperature. However, this is impossible for most of the events since the Solar Orbiter / SWA instrument  was not observing continuously. For the detected whistler waves with available plasma measurements, during the time interval between July and October 2020, the solar wind velocity was found to be between 260 km/s and 400 km/s about 88\% of the time, and never larger than ~530 km/s. This signifies that the results presented here may be considered to be representative of a slow to intermediate winds, but  not of a fast wind.  Furthermore, for the first part of the orbit between February and May 2020, the WIND measurements at L1 show that the solar wind velocity were most of the time below 450 km/s with a few and short incursions above 500 km/s . The fewer number (see below) of whistler waves observed during this period, when Solar Orbiter was not too far from the Earth, then does not seem to be caused by the presence of a faster wind, and the dependence on distance that we observed appears to be real.

\subsection{Occurrence rate}
  \begin{figure}
   \centering
    \includegraphics[trim={0cm 1cm 0cm 2cm},width=0.49\textwidth]{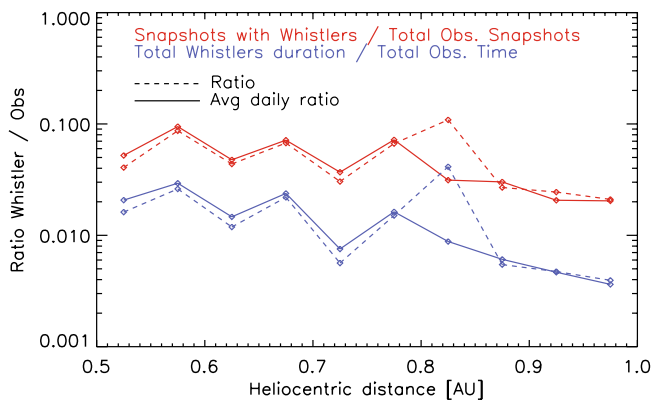}
   \caption{Variations of whistler occurrence with heliocentric distance. The numbers of snapshots with whistler waves for a given heliocentric distance bin ranges from 237 to 1050.}
              \label{FigOccurrence}%
    \end{figure}
\begin{figure*}
\centering
   \includegraphics[trim={4cm 2cm 0cm 2cm},width=0.49\textwidth]{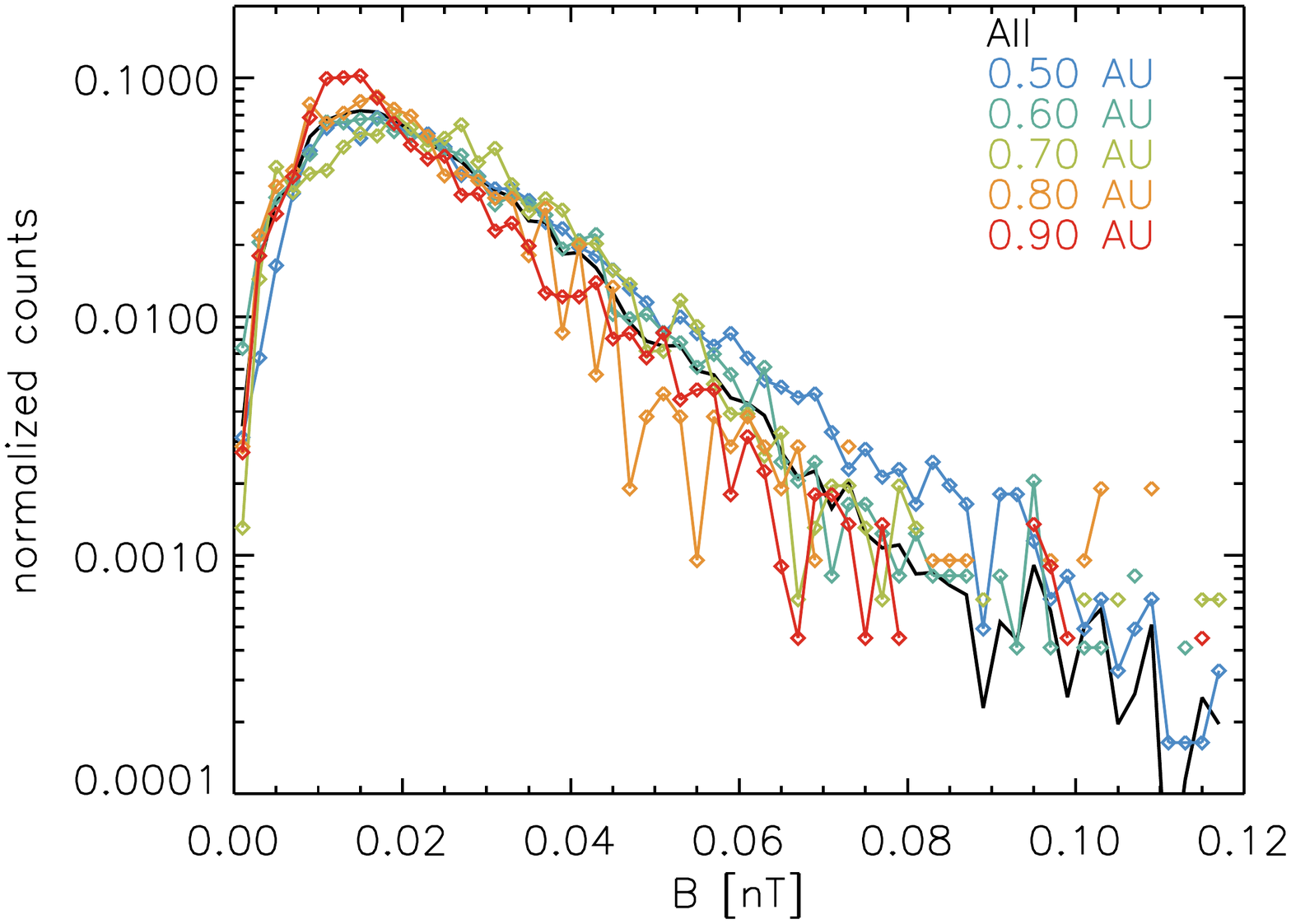}\includegraphics[trim={4cm 2cm 0cm 2cm},width=0.49\textwidth]{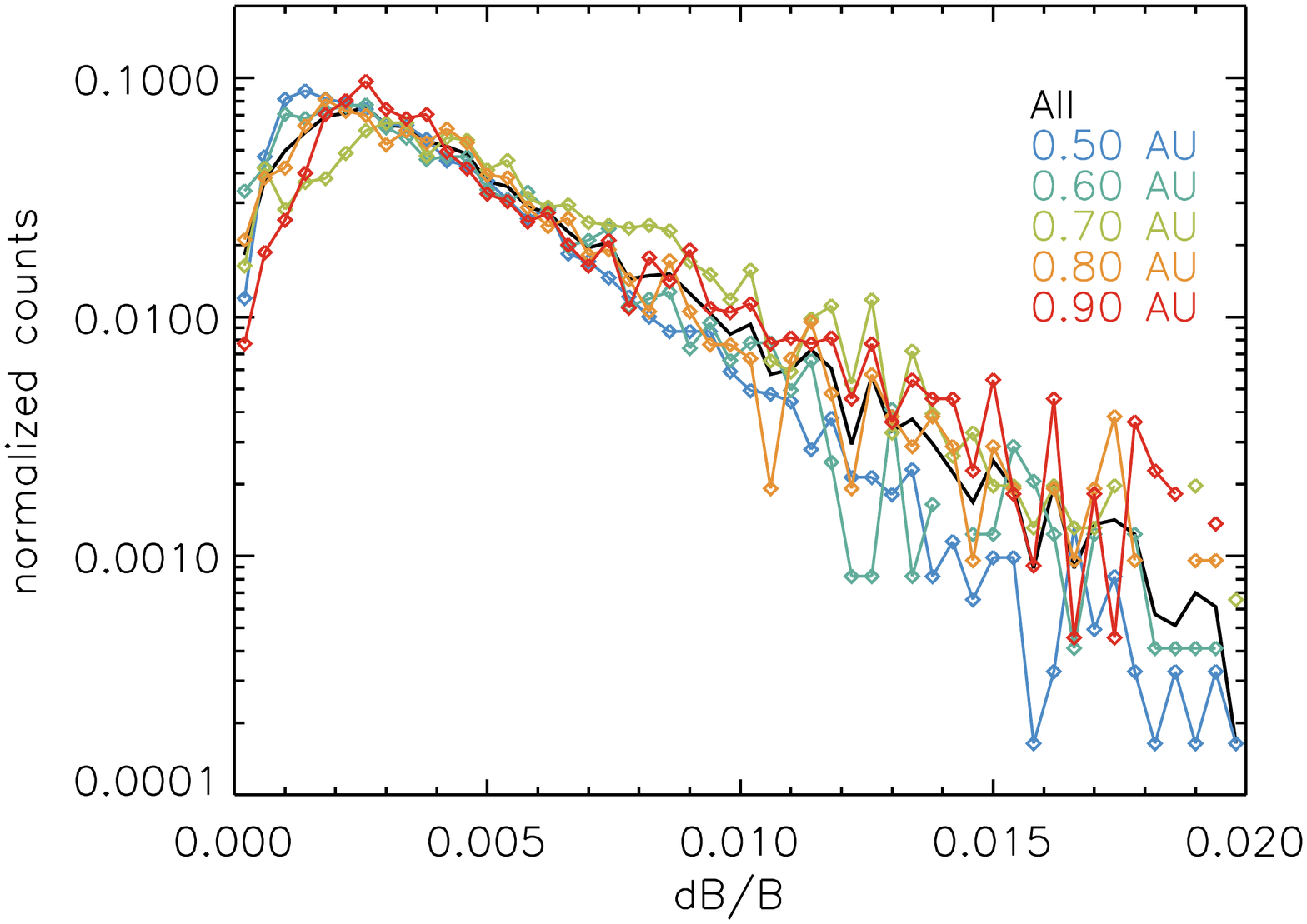}
   \includegraphics[trim={4cm 2cm 0cm 2cm},width=0.49\textwidth]{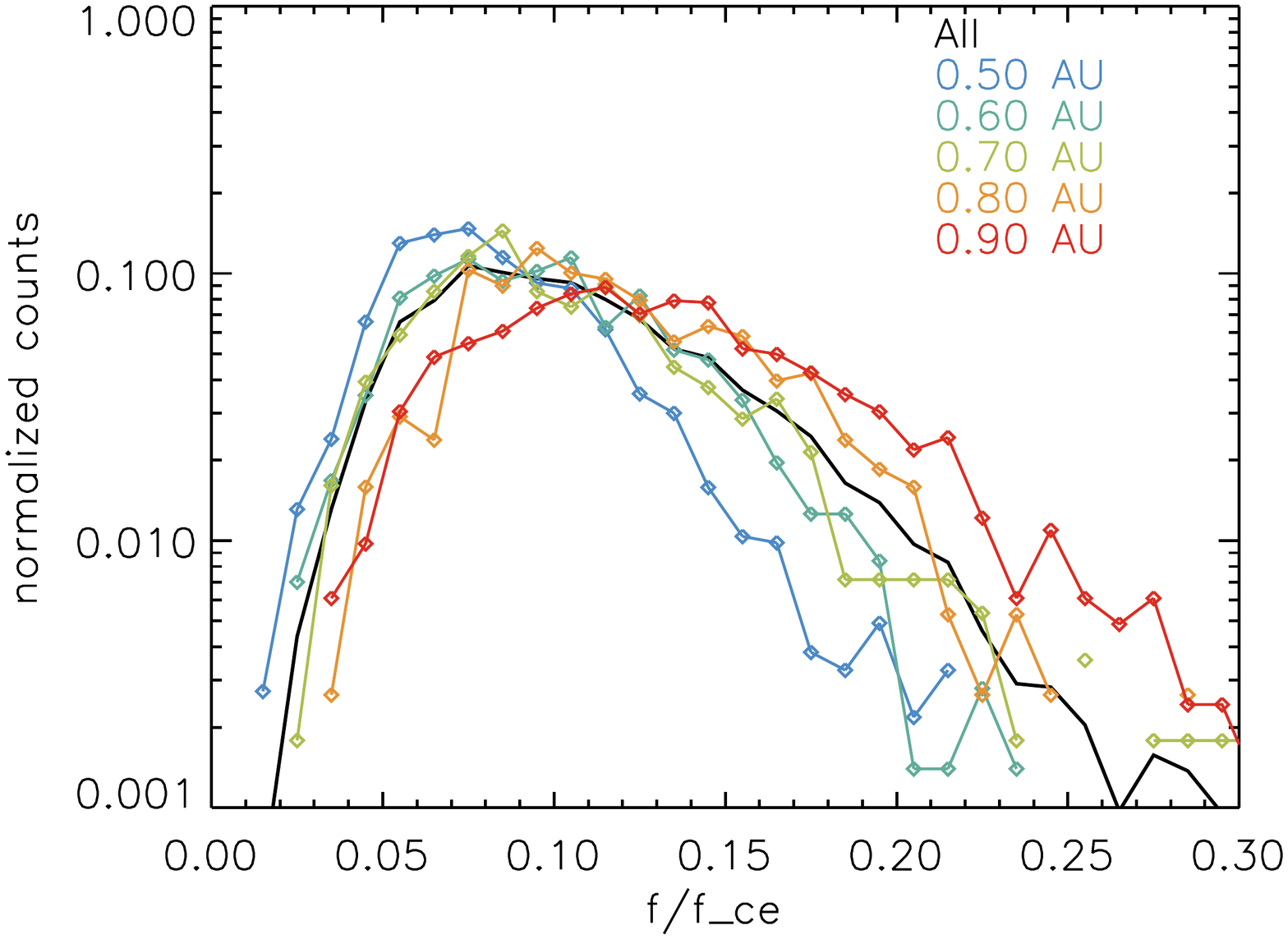}\includegraphics[trim={4cm 2cm 0cm 2cm},width=0.49\textwidth]{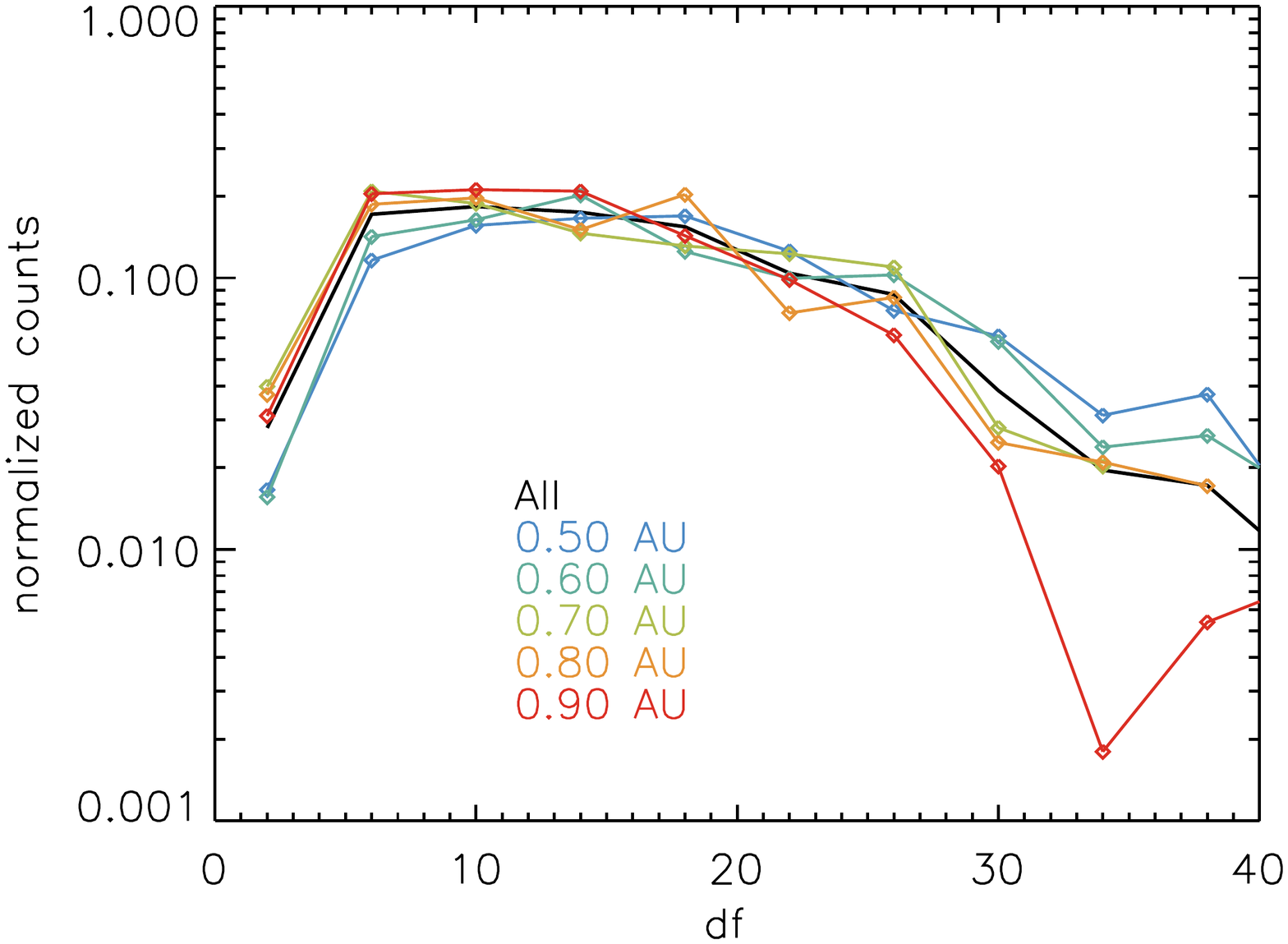}
     \caption{Normalized histogram for various wave parameters and different heliocentric distances. \textit{Top left}: Wave magnetic amplitude  \textit{Top right}: Wave magnetic amplitude normalized to background field. \textit{Bottom left}: Normalized frequency in solar wind frame. \textit{Bottom right:} frequency width. }
     \label{FigHisto}
\end{figure*}

We have estimated the occurrence rate of whistler waves in two ways. First, we evaluate the ratio of the number of snapshots with at least one whistler wave packet to the total number of observed snapshots, as was done in previous studies \citep{2019ApJ...878...41T,2020ApJ...897..118J,2021arXiv210106723J}. This provides us with an upper estimate of the occurrence rate, as it is unlikely that the wave packet stands for the  whole duration of the snapshot. Second, we have computed the ratio between the total observed time including whistler waves, as the sum of the duration of each wave packet, and the total time of observations. This provides us with a more precise estimate of their occurrence rate. Both estimates are however limited by the sensitivity of SCM. Fig. \ref{FigOccurrence} shows the occurrence rate, computed over each heliocentric bin and as averaged daily value within the bin (daily ratio averaged over the heliocentric bins). The value at 0.825 AU that corresponds to observations between 0.8 and 0.85 AU differs for the two computations. This can be explained by the presence in that bin of the magnetic perturbation that crossed Solar Orbiter on 13 April 2020. On that day, nearly 4000 whistler wave packets have been detected, which explains why the averaged daily ratio is smaller than the ratio computed over the whole interval (19 days). This also evidences that the whistler mode occurrence is not steady and depends on wind conditions, as has been shown by the previous studies. \\
The value obtained in the bin [0.95-1AU] with the 8 s average spectrum (snapshot ratio) is close to the $\sim$ 1.7\%  occurrence rate observed at 1AU by \textit{Artemis} \citep{2019ApJ...878...41T}. It is however about 10 times less than the value obtained by\cite{2020ApJ...897..118J} with  HELIOS data. \\
In despite of the statistical fluctuations, one can notice that there is a general trend for whistler waves to occur more often when going from 1 AU to 0.5 AU, varying from an occurrence of $\sim $2\% near the Earth to  5-10\% at 0.5 AU (using the snapshot ratio). This is in contradiction with the results of \cite{2020ApJ...897..118J}, who analyzed HELIOS data and found the occurrence to decrease from $\sim $15\% to $\sim $3\%. 
They had only low resolution frequency spectra at their disposal and their selection criteria were therefore based only on bump in these spectra. It is therefore possible that Doppler shifted ion cyclotron waves were counted as whistler waves, although our observations do not support an important presence of ICW at these wavelengths. Another possible explanation resides in the fact that the Helios search coil magnetometer had longer antennas and a sensitivity about 5 times better at low frequencies than SCM on Solar Orbiter. \cite{2020ApJ...897..118J} reported mean wave amplitude between 0.01 nT and 0.02 nT at 1AU, and around 0.03 nT - 0.04 nT at 0.5 AU. We found the mean amplitude to increase from 0.02 nT to 0.026 nT, therefore significantly less intense whistler waves closer to the Sun but similar values at 1AU (the distribution of the wave amplitudes is shown in the top left panel of fig.\ref{FigHisto}). We measured amplitudes down to 10$^{-3}$ nT and 18\% of the waves at 1 AU have amplitudes below 0.01 nT. Visual inspection of  the distribution does not indicate that we are missing small values neither. Therefore, it looks improbable that this difference in sensitivity can explain the difference in the observed occurrence. Another difference between the two studies is that we did not exclude planetary shocks and magnetic clouds, to the noticeable exception of the event on April 13. \cite{2020ApJ...897..118J} found the occurrence rate of whistler waves to depend on solar wind velocity, with the largest rates for the slowest wind. However, their slow wind observations also correspond to distance below 0.5 AU. \cite{2019ApJ...878...41T} did not find any dependence of the occurrence rate on solar wind velocity between 270 km/s and 630 km/s.
The fewer number of whistler waves at larger distances from the Sun that we observed here, corresponds in part to the spring of 2020 when Solar Orbiter was not too far from the Earth and when WIND observations indicate that slow wind was dominant. Therefore, this does not seem to be related to the presence of a faster wind far from the Sun; unless the wind speed was slower enough at 0.5 AU to compensate for an increasing rate with shorter distance, our finding of more numerous whistler waves at 0.5 AU than at 1AU appears in contradiction with Helios observation.

\subsection{Wave properties}

In this section we present the statistical characteristics of the observed whistler waves and their variations with the heliocentric distance. We consider the perturbation that occurred on April 13, 2020 as a sufficiently exceptional event to exclude the waves observed on that day in the following. \\
The top panels of fig.\ref{FigHisto} show the distribution of the wave's amplitudes at various distances. Whistler waves occurring at closer distance to the Sun have larger amplitudes, which may be interpreted as reflecting a larger amount of free energy available for the generation of waves, in the electron heat flux or in electron anisotropy, for example, at this distance. Although \cite{2020ApJ...897..118J} reported no clear trend of the wave amplitude with the heliocentric distance, their figure shows more intense whistler waves closer to the Sun, which agrees with our results. When normalized to the background magnetic field, the wave amplitudes appear to be slightly smaller at 0.5 AU than at larger distance, showing that the wave amplitudes do not scale with the background magnetic field but depend on other parameters.  

The bottom panels of fig.\ref{FigHisto} show the histograms of the frequency and frequency bandwidth for different heliocentric distances. Whistler waves occurring closer to the Sun have lower maximum frequency but larger frequency bandwidth than at 1AU. This larger frequency bandwidth implies larger energy content in the waves and thus can also be interpreted as reflecting a larger amount of available energy. 

\begin{figure}
   \centering
    \includegraphics[trim={4.5cm 1.5cm 1.5cm 2cm},width=0.4\textwidth]{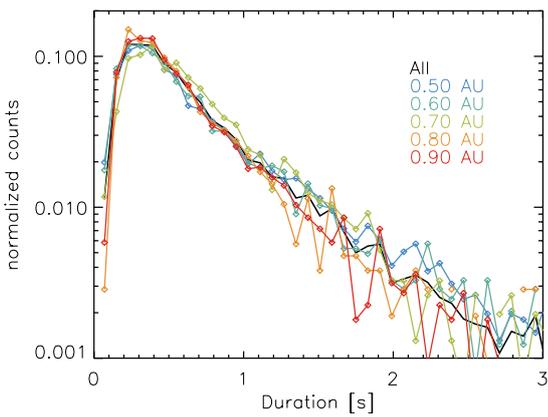}
    \caption{Distribution of wave packet duration for various heliocentric distances}
              \label{FigDuration}%
    \end{figure}

Fig.\ref{FigDuration} shows a tendency for wave packets to be longer at smaller distances from the Sun. This could, however, be an effect of larger wave amplitude that would make the wave packet standing above the noise level for longer time.  

\begin{figure}
   \centering
     \includegraphics[trim={4.5cm 1.5cm 1.5cm 2cm},width=0.4\textwidth]{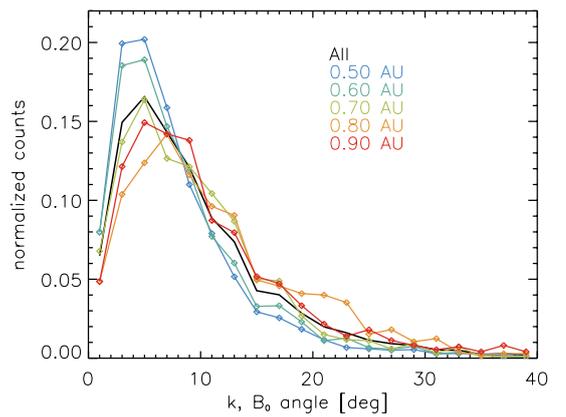}
    \caption{Distribution of the k-vector angles with the background magnetic field}
              \label{FigTheta}%
    \end{figure}

Fig.\ref{FigTheta} shows for various heliocentric distances the normalized histogram of the angle between the $\mathbf{k}$ vector and the background magnetic field $\mathbf{B_0}$. As shown previously, most of the waves are quasi parallel, with angles less than 20 $^\circ$. We can also notice that waves observed near 0.5 AU are more aligned than at 1 AU, a trend that is clear but for which we have no clear explanation for now.   \\






\section{Conclusion}\label{sec_conclu}
In this paper, we have reported the results of the study of the electromagnetic waves observed by Solar Orbiter in the frequency range between 3 Hz and 128 Hz during its first orbit, covering heliocentric distances between 0.5 AU and 1 AU. Using electric and magnetic fields measurements provided by the RPW instrument, we have shown that an overwhelming majority of these waves are right-hand circularly polarized in the solar wind frame and correspond to whistler waves. They propagate  quasi parallel to the background magnetic field and anti sunward. This is consistent with  the whistler heat flux instability scenario for the regulation of the heat flux. It is worth noting that this study does not rule out the presence of oblique whistler waves, as these later may be quasi electrostatic and therefore more difficult to detect with our selection procedures based on magnetic measurements. \\
Contrary to what was found using HELIOS observations, we observed more whistler waves closer to the Sun, by at least a factor of 2. Solar Orbiter observations during the next orbits will improve the statistics over these distances. \\
Whistler waves are generally intermittent and clumped in short duration ($\sim 0.3$ s) wave packets. When taking into account their actual duration, their occurrence varies from 0.3\% at 1 AU to 1\%-2\% at 0.5 AU. Whistler wave packets are presumably slightly longer at 0.5 AU than at 1AU, although this could be an effect of their detectability as they also have larger wave power at 0.5 AU. \\
We also found that whistler waves tend to be more field-aligned and to have smaller normalized frequency ($f/f_{ce}$), larger amplitude, and larger bandwidth at 0.5 AU than at 1AU. The larger occurrence, amplitude, and bandwidth clearly indicates that the source of energy that creates whistler waves increases while approaching the Sun from 1 AU to 0.5 AU.
%

%
 
%

\begin{acknowledgements}
 Solar Orbiter is a space mission of international collaboration between ESA and NASA, operated by ESA. We thank the Centre National d’Etudes Spatiales (CNES, french space agency) for having funded the Search Coil magnetometer and several subsystems of RPW. DG and YK is supported by the Swedish National Space Agency grant 20/136. Solar Orbiter magnetometer operations are funded by the UK Space Agency (grant ST/T001062/1). Tim Horbury is supported by STFC grant ST/S000364/1
\end{acknowledgements}
\bibliographystyle{aa}
\bibliography{MyBibFromPapers_last.bib}

%
%
\newpage

\newpage

\end{document}